# Phase transitions and self-assemblies
# of lower diamondoids and derivatives.

Yong Xue [1], G.Ali Mansoori [2]
University of Illinois at Chicago, (M/C 063) Chicago, IL 60607-7052, USA,
(1). xueyong37@gmail.com; (2). mansoori@uic.edu

**ABSTRACT**

Applying *ab initio* calculation and molecular dynamics simulation methods, we have been calculating and predicting the essential self-assemblies and phase transitions of two lower diamondoids (adamantane and diamantane), three of their important derivatives (amantadine, memantine and rimantadine), and two organometallic molecules that are built by substituting one hydrogen ion with one sodium ion in both adamantane and diamantine molecules (ADM•Na and Optimized DIM•Na). To study their self-assembly and phase transition behaviors, we built seven different MD simulation systems, and each system consisting of 125 molecules. We obtained self-assembly structures and simulation trajectories for the seven molecules. Radial distribution function studies showed clear phase transitions for the seven molecules. Higher aggregation temperatures were observed for diamondoid derivatives. We also studied the density dependence of the phase transition which demonstrates that the higher the density - the higher the phase transition points.

**INTRODUCTION**

Diamondoids and their derivatives have found major applications as templates and as molecular building blocks in nanotechnology [1-3]. They have been drawn more and more researchers' attentions to their highly symmetrical and strain free structures, controllable nanostructural characteristics, non-toxicity and their applications in producing variety of nanostructure shapes, in molecular manufacturing, in nanotechnology and in NEMS and MEMS. It is important and necessary to study self-assembly of these molecules in order to obtain reference data, such as temperature, pressure, bonding properties, etc. for application in nanotechnology e.g. building molecular electronic devices.

**THEORY**

Two lower diamondoids, three adamantane derivatives and two artificial molecules (substituting one hydrogen atom in adamantane and diamantine with one sodium atom) are studied in this report. We classified them into three groups as shown in Table I.
Group 1: Adamantane (ADM) and Diamantane (DIM), the lowest two diamondoids. Due to their six or more linking groups, they have found major applications as templates and as molecular building blocks in nanotechnology, polymer synthesis, drug delivery, drug targeting, DNA-directed assembly, DNA-amino acid nanostructure formation, and host-guest chemistry [1-3]. However these diamondoids do not have good electronic properties which are necessary for building molecular electronics, but some of their derivatives do. Group 2: Memantine, Rimantadine and Amantadine, the three derivatives of adamantane, which have medical applications as antiviral agents and due to their amino groups, they could be treated as molecular semiconductors [4]. Group 3: ADM•Na, DIM•Na, the two organometallic molecules (substituting one hydrogen ion in adamantane and diamantane with a sodium ion) which could have potential applications in NEMS and MEMS [2]. This report is aimed at studying the self-assembly and phase transition properties of these seven molecules for further understanding of their structures and possibly building molecular electronic devices with them.

**Table I.** Molecular formulas and structures of Adamantane, Diamantane, Amantadine, Rimantadine, Memantine, Optimized ADM•Na and Optimized DIM•Na molecules. In these figures blacks represent –C, whites represent –H, Blues represent –N and purples represent –Na.

| Group 1 | | Group 2 | | | Group 3 | |
|---|---|---|---|---|---|---|
| Adamantane | Diamantane | Amantadine | Rimantadine | Memantine | Optimized ADM•Na | Optimized DIM•Na |
| $C_{10}H_{16}$ | $C_{14}H_{20}$ | $C_{10}H_{17}N$ | $C_{11}H_{20}N$ | $C_{12}H_{21}N$ | $C_{10}H_{15}Na$ | $C_{14}H_{19}Na$ |
| 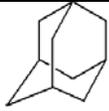 | 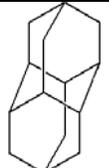 | 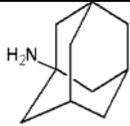 | 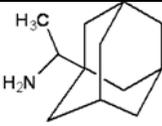 | 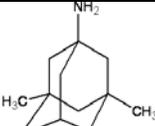 | 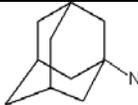 | 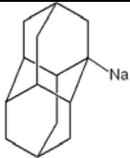 |

We first performed density functional theory (DFT) calculations to optimize initial geometric structures of theses seven molecules and we obtained atomic electronic charges. Then we performed molecular dynamics (MD) simulation for the study of their self-assembly and phase transition behaviors [1].

**DISCUSSION**

**Self-assembly of lower diamondoids**

In order to find the equilibrium configurations of the collection of the 125 molecules at every given temperature we used the simulated-annealing procedure. Every 1 K temperature change occurred within 10 picoseconds (5,000 time-steps were used and each time-step was 0.002 ps). With these settings we studied the self-assembly behaviors in the cooling steps [1].

The self-assembly snapshots of the MD simulations for 125 molecules of each of the seven molecules are reported in Figure 1. From these snapshots we can directly observe clear phase transitions for each kind of molecule, from the gaseous state to their aggregation into a highly condensed solid (self-assembled) state. Group 1 and group 3 showed comparatively decent crystalline structures, while Group 2 demonstrated non-ordered structures.

The final simulation structure of adamantane and the corresponding temperature matched the experimental data of adamantane at solid state phase in general. These facts could be validations of our simulation methods. [1]

The -NH$_2$ and -CH$_3$ residues raise the phase transition temperatures of derivatives comparing adamantane molecules; at the same time, however, these extra configurations violent the symmetrical structure of single adamantane molecules and break the final neat crystal structures. While Van Der Waal bonding played a central role in these self-assembly and phase transition procedures, hydrogen bonds were observed as well in Group 2, elaborated in next section.

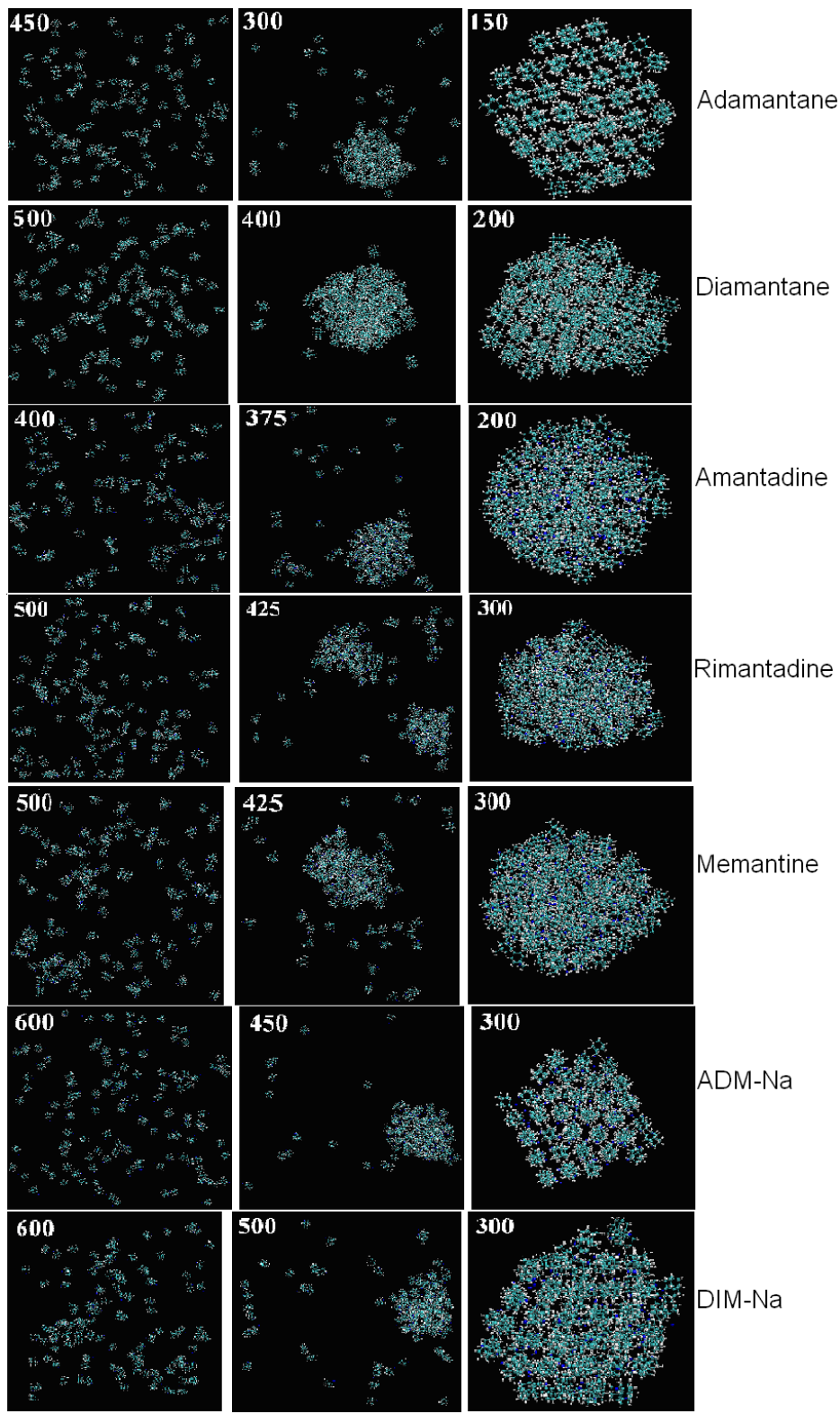

**Figure 1**. Various stages of self-assembly process snapshots of 125 molecules of the seven compounds as temperatures [K] decrease. From left to right: vapor, an intermediate self-assembly stage, completed self-assembly (solid).

**Hydrogen-bonds distribution study**

We further investigated the hydrogen-bonds distributions of the self-assembled snapshots of amantadine, memantine and rimantadine at 50K as shown in Figure 2. According to this figure there is no ordered structural pattern for the location of hydrogen-bonds even at adequately low temperature which is an indication of the non-crystalline self-assembled states of these three molecules. This observation also manifested that the -$NH_2$ and -$CH_3$ residues prevented the formations of clear crystalline state due to their non-symmetrical configurations, while -$NH_2$ contributed to the hydrogen bonds which in return alleviated the phase transition temperatures.

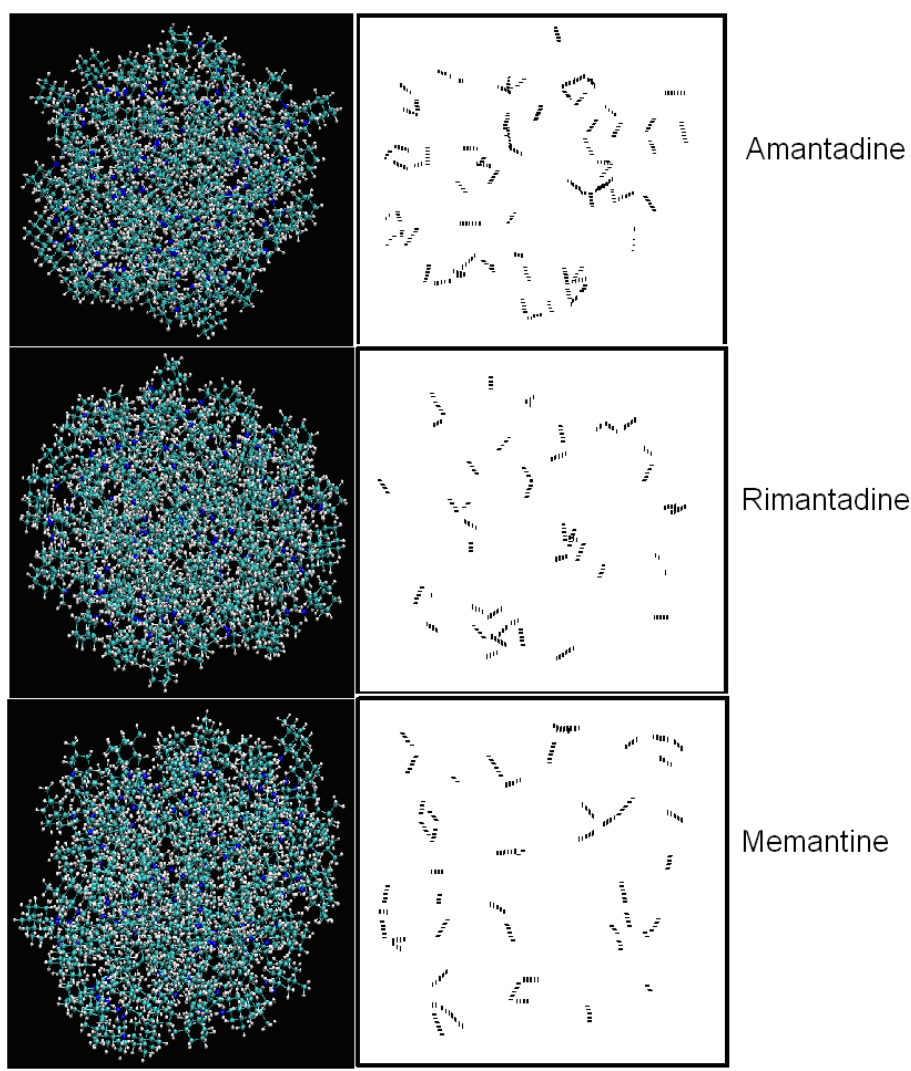

**Figure 2**. MD snapshots of Group 2 and their hydrogen bonds locations at 50K. The randomness of the hydrogen-bond distributions indicates the non-crystalline structures of this group.

## Radial Distribution Functions (RDF) study

In Figures 3 we report the self-assembled (solid-state) radial distribution functions of the seven molecules with uniform coordinates scales for the purpose of their collective comparison.

From this figure we can observe that adamantane and adamantane+Na RDFs have more sharp peaks than RDFs of the other five molecules. This indicates distinct crystalline solid state structures for adamantane and adamantane+Na. While the other five molecules also show solid characteristic peaks, they have less number of such sharp peaks than adamantane and adamantane+Na. These results match the image observation of simulations (Figure 1), i.e. adamantane and adamantane+Na have neater self-assembly structures in solid state.

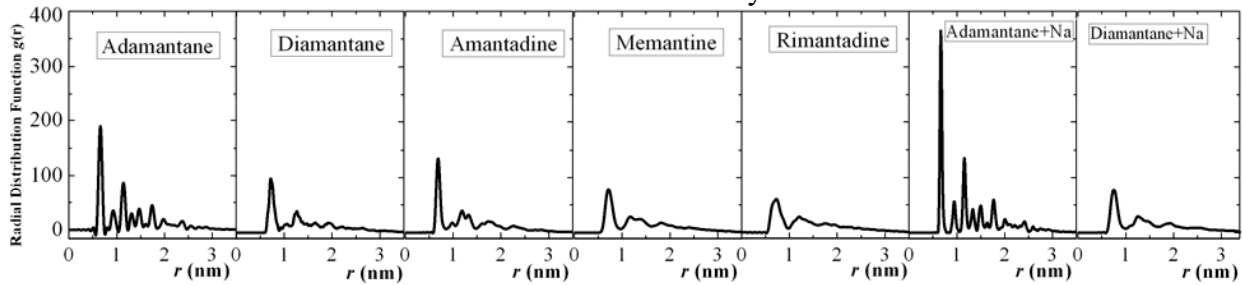

**Figure 3**. Radial distribution functions of the seven compounds (from left: Adamantane, Diamantane, Amantadine, Rimantadine, Memantine, Adamantane+Na., Diamantane+Na) in the self-assembled (solid) state.

## Density dependence studies of phase transition points

We chose different volumes for 64 and 125 adamantane molecular systems to study the density dependence of phase transition points (Onset and completion of self-assembly) as shown in Figure 4 and Figure 5. It is obvious that transition points vary when the density changes for both 64 and 125 adamantane molecular systems. The transition points, however, are markedly different at low densities for the two systems of 64 and 125 molecules.

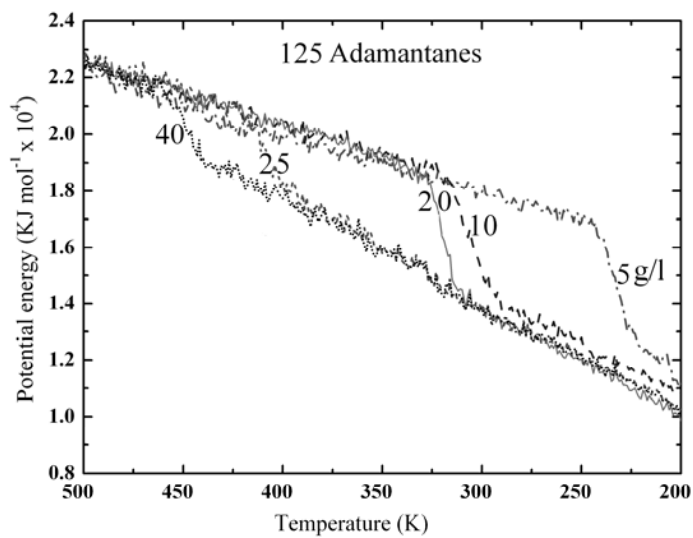

**Figure 4**. Potential energy vs. Temperature raw simulation data of 125 adamantane molecules ensemble system at different density from 5, 10, 20, 25 and 40g/l simulated annealing from 500K-200K, showing density dependence of phase transition points.

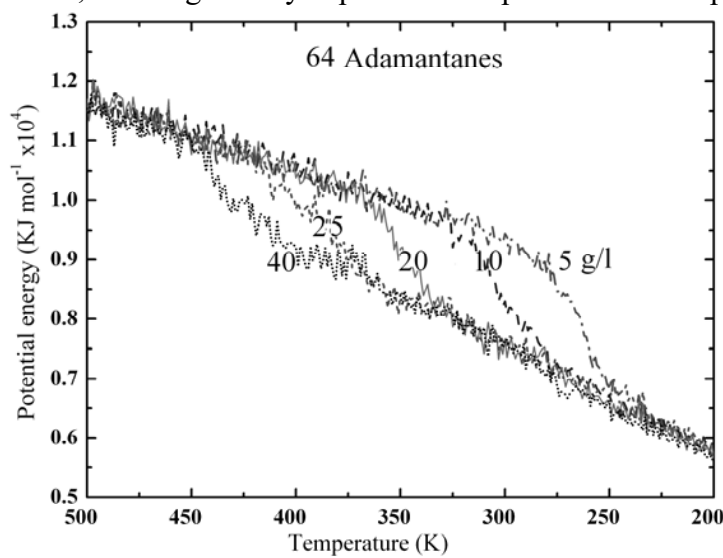

**Figure 5**. Total potential energy vs. Temperature raw simulation data of 64 adamantane molecules ensemble system at different densities (5, 10, 20, 25 and 40g/l) using simulated annealing technique showing density dependence of phase transition points.

**CONCLUSIONS**

Our results indicate: 1. The nature of self-assembly and phase transition in these molecules is a structure-dependent phenomenon. 2. The final self-assemblies possess crystalline structures when no hydrogen-bonding is present in the molecular structure. 3. The organometallic molecules also hold neat crystal structures. Although -Na ion increases the phase transition temperature, as those -$NH_2$ ions in group 2. To a large extent the crystalline structural features of diamondoids are retained in adamantane-Na and diamantane-Na self-assemblies. The reasons for the latter might be that: A. The –Na ion has less topology effect than does the –$NH_2$ ion. B. There is no hydrogen-bonding in the structures of adamantane-Na and diamantane-Na; therefore they can aggregate to ordered structures. This feature is very promising, since it allows us to build orderly-shaped NEMS and MEMS. The density dependence of phase transition point could assist the controlling of the self-assembly processes of those molecules.